# Up to Nine Millennia of Multimessenger Solar Activity

Leif Svalgaard, Stanford University (leif@leif.org)

**Abstract**

A nine-millennia reconstruction of decadal sunspot numbers derived from $^{10}$Be and $^{14}$C terrestrial archives for 6755 BC to 1885 AD has been extended to the present using several other messengers (Observed Sunspot Number, Group Number, range of the diurnal variation of the geomagnetic field, and the InterDiurnal Variation of the geomagnetic Ring Current) and scaled to the modern SILSO Version 2 sunspot number. We find that there has been no secular up tick of activity the last three hundred years and that recent activity has not been out of the ordinary. There is a sharp 87.6-year peak in the power spectrum, but no significant power at the Hallstatt 2300-year period. The reconciliation of the cosmogenic record with the modern sunspot record could be an important step to providing a vetted solar activity record for the use in climate research.

**Introduction**

Wu et al. (2018) (hereafter WEA) present a multi-proxy reconstruction of solar activity over the last 9000 years, using all available long-span datasets of $^{10}$Be and $^{14}$C messengers in terrestrial archives. Cosmogenic isotopes are produced by cosmic rays in the Earth's atmosphere and their measured production/depositional flux reflects changes in the cosmic ray flux in the past. The cosmic ray flux is modulated by solar magnetic activity, which can be quantified in terms of the heliospheric modulation potential characterizing the energy spectrum of Galactic Cosmic Rays reaching the top of the atmosphere at a given time. The WEA reconstruction is given as decadal averages centered on the midpoint of each decade and runs from 6755.5 BC to 1885.5 AD. The reason for stopping in 1885 was that the (Suess 1955) effect of extensive fossil fuel burning makes it problematic to use $^{14}$C data after the mid-19$^{th}$ century; in addition, radiocarbon data cannot be used after the 1950s because of nuclear explosions that led to massive production of $^{14}$C. The modulation potential series is not a stable proxy for solar activity since the modulation potential is a relative index whose absolute value is model dependent (e.g. Herbst et al. 2017). Therefore, WEA converted the reconstructed modulation potential to a more practical and certainly more widely used index: the sunspot number, its current version designated SN (version 2, Clette et al. 2014). The conversion was done via the open solar magnetic flux following an 'established' procedure (e.g. Usoskin et al. 2003, 2007). As the 'procedure' was developed for version 1 of the sunspot number, the newer version 2 data were scaled down by a factor of 0.6 for the calibration, in spite of the so-called k-factor (the 0.6) not being constant over time (Clette & Lefèvre 2016). It seems a step backwards to cling to the obsolete version 1 of the sunspot number scale, so we undo the spurious down-scaling of version 2. We shall not here quibble about details of the conversion procedure except to note that one would expect (even require) that the SN-reconstruction should match the actual observed SN-series for the time of overlap. WEA suggest that their reconstructed values be multiplied by 1.667 to place them on the SN V2-scale. Figure 1 shows that this is not enough. A factor of 2.0 seems to be necessary to match the two scales, likely meaning that the WEA calibration is too low by about 20%.

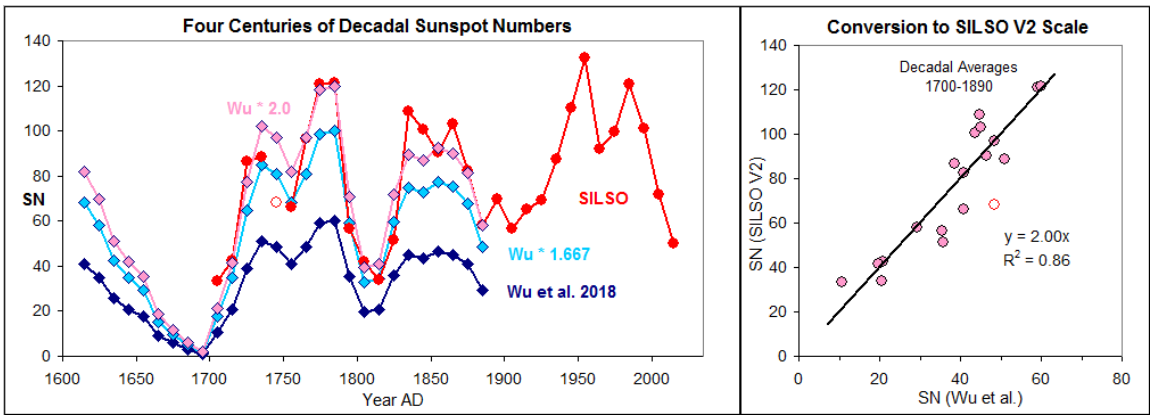

Figure 1. (Left) Wu et al. (2018) reconstructed decadal sunspot numbers (dark blue diamonds) since 1600 AD. Undoing the down-scaling, yields the curve with light blue diamonds, but this is not enough to match the SILSO SN version 2 (red dots). (Right) A factor of 2.0±0.2 seems to be needed. The SN value for the 1745 decade is treated as an outlier (open circle) as there were no observations for several years during the decade.

**More Messengers**

When comparing two series it can be difficult to decide which one is too low or too high. It could simply be wrong. Luckily, there are several other messengers directly pertaining to solar activity: the independently derived sunspot Group Number, GN (Svalgaard & Schatten 2016) back to 1610, the range of the diurnal variation of the geomagnetic field, rY (Svalgaard 2016; Loomis 1873) good back to the 1810s, and the InterDiurnal Variation of the geomagnetic Ring Current, IDV (Svalgaard & Cliver 2005, 2010; Svalgaard 2014; Cliver & Herbst 2018; Owen et al. 2016; Bartels 1932) back to the 1830s. Decadal means for these are given in Table 1 together with the (linear) regression-equations to convert them to the SN V2 scale. Applying the conversions we can now plot the messages all on the same scale, Figure 2.

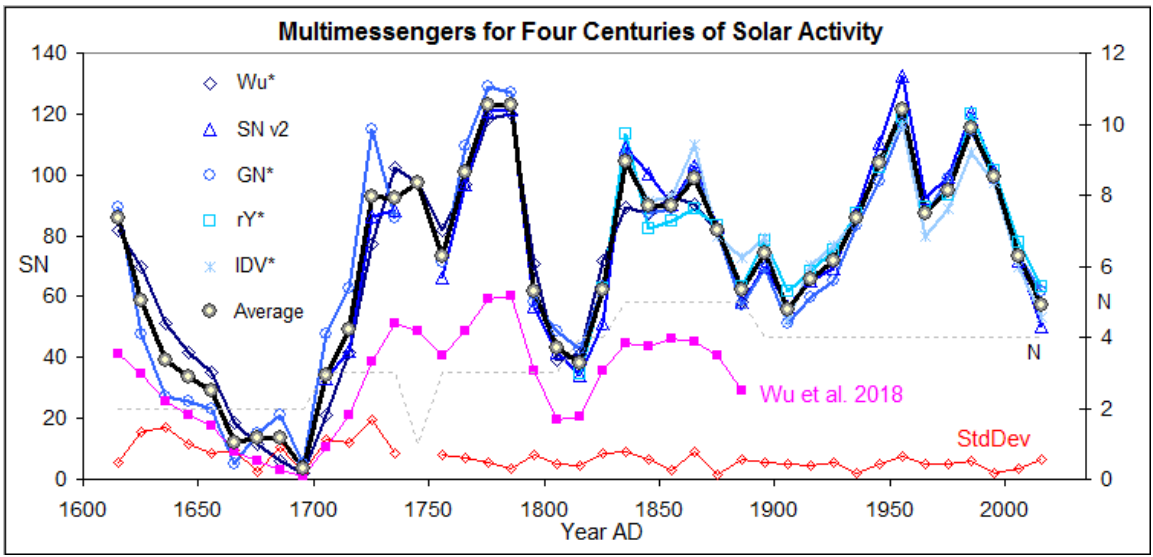

Figure 2. Four centuries of reconstructed decadal sunspot numbers on the SILSO V2 scale. We have direct observations of sunspot numbers (bluish triangles), of sunspot group numbers (GN* bluish circles; the asterisk * signifies reconstructed values using the regression coefficients given in Table 1 below), of daily variation ranges (rY* bluish squares), of ring current strength (IDV* bluish stars), and of the corrected cosmogenic series (Wu* blue diamonds scaled up by a factor of 2, as per Figure 1; the original series is shown by the pink squares in the middle of the graph). Gray dots with a yellow center show the average of all the messages (with standard deviations shown by red diamonds at the bottom of the Figure). The number of values, N, in each average is marked by the thin dashed gray curve.

In scaling rY and IDV we have first constructed a composite of SN V2 and GN*(on SN V2 scale).

Table 1. The published values of the decadal averages for the five messengers considered in this article are tabulated in columns 2-6. Columns 7-11 contain the their messages put on the SILSO SN Version 2 scale using the regression coefficients given in rows 2 and 3, e.g. IDV*(V2) = 18.71*IDV(nT) – 91.27. Column 12 gives the average of columns 7-11, with its standard deviation in column 13, based on the number of values, N in column 14, going into the average. The table can also be found in the Excel file (see below) associated with this article.

| | | | | Decadal Sunspot Numbers and Proxies | | | | | | | | | |
|---|---|---|---|---|---|---|---|---|---|---|---|---|---|
| | reconstr | | | | | 2.00 | 1 | 18.65 | 7.63 | 18.71 | | | |
| | Wu ea. | SILSO | S&S | Svalgaard | Svalgaard | | | | -238 | -91.27 | | | |
| Year AD | SNN v1 | SN v2 | GN | rY nT | IDV nT | SNN* | SN v2 | GN S&S* | rY* | IDV* | Avg SN | Std Dev | N |
| 1615.5 | 40.907 | | 4.80 | | | 81.8 | | 89.5 | | | 86 | 5.4 | 2 |
| 1625.5 | 34.847 | | 2.56 | | | 69.7 | | 47.7 | | | 59 | 15.5 | 2 |
| 1635.5 | 25.608 | | 1.45 | | | 51.2 | | 27.0 | | | 39 | 17.1 | 2 |
| 1645.5 | 20.939 | | 1.36 | | | 41.9 | | 25.4 | | | 34 | 11.7 | 2 |
| 1655.5 | 17.667 | | 1.25 | | | 35.3 | | 23.3 | | | 29 | 8.5 | 2 |
| 1665.5 | 9.239 | | 0.28 | | | 18.5 | | 5.2 | | | 12 | 9.4 | 2 |
| 1675.5 | 5.867 | | 0.81 | | | 11.7 | | 15.1 | | | 13 | 2.4 | 2 |
| 1685.5 | 3.016 | | 1.13 | | | 6.0 | | 21.1 | | | 14 | 10.6 | 2 |
| 1695.5 | 1.035 | | 0.28 | | | 2.1 | | 5.2 | | | 4 | 2.2 | 2 |
| 1705.5 | 10.663 | 33.2 | 2.56 | | | 21.3 | 33.2 | 47.7 | | | 34 | 13.2 | 3 |
| 1715.5 | 20.839 | 42.4 | 3.37 | | | 41.7 | 42.4 | 62.9 | | | 49 | 12.0 | 3 |
| 1725.5 | 38.700 | 86.5 | 6.16 | | | 77.4 | 86.5 | 114.9 | | | 93 | 19.6 | 3 |
| 1735.5 | 51.083 | 88.5 | 4.61 | | | 102.2 | 88.5 | 86.0 | | | 92 | 8.7 | 3 |
| 1745.5 | 48.559 | | | | | 97.1 | | | | | 97 | | 1 |
| 1755.5 | 40.873 | 66.4 | 3.83 | | | 81.7 | 66.4 | 71.4 | | | 73 | 7.8 | 3 |
| 1765.5 | 48.425 | 96.7 | 5.86 | | | 96.9 | 96.7 | 109.3 | | | 101 | 7.2 | 3 |
| 1775.5 | 59.153 | 121.0 | 6.91 | | | 118.3 | 121.0 | 128.9 | | | 123 | 5.5 | 3 |
| 1785.5 | 60.002 | 121.3 | 6.80 | | | 120.0 | 121.3 | 126.8 | | | 123 | 3.6 | 3 |
| 1795.5 | 35.432 | 56.6 | 3.12 | | | 70.9 | 56.6 | 58.2 | | | 62 | 7.8 | 3 |
| 1805.5 | 19.676 | 41.7 | 2.61 | | | 39.4 | 41.7 | 48.7 | | | 43 | 4.8 | 3 |
| 1815.5 | 20.548 | 33.9 | 2.30 | 35.70 | | 41.1 | 33.9 | 42.9 | 34.4 | | 38 | 4.6 | 4 |
| 1825.5 | 35.815 | 51.4 | 3.39 | 39.44 | | 71.6 | 51.4 | 63.2 | 62.9 | | 62 | 8.3 | 4 |
| 1835.5 | 44.732 | 108.9 | 5.69 | 46.06 | 10.5 | 89.5 | 108.9 | 106.1 | 113.4 | 105.2 | 105 | 9.1 | 5 |
| 1845.5 | 43.531 | 100.4 | 4.68 | 42.00 | 9.77 | 87.1 | 100.4 | 87.3 | 82.5 | 91.5 | 90 | 6.8 | 5 |
| 1855.5 | 46.352 | 90.3 | 4.77 | 42.32 | 9.83 | 92.7 | 90.3 | 89.0 | 84.9 | 92.6 | 90 | 3.2 | 5 |
| 1865.5 | 45.088 | 103.1 | 5.47 | 42.84 | 10.74 | 90.2 | 103.1 | 102.0 | 88.9 | 109.7 | 99 | 8.9 | 5 |
| 1875.5 | 40.713 | 82.6 | 4.36 | 42.14 | 9.15 | 81.4 | 82.6 | 81.3 | 83.5 | 79.9 | 82 | 1.4 | 5 |
| 1885.5 | 29.175 | 58.1 | 3.07 | 39.50 | 8.78 | 58.4 | 58.1 | 57.3 | 63.4 | 73.0 | 62 | 6.6 | 5 |
| 1895.5 | | 69.6 | 3.72 | 41.44 | 9.11 | | 69.6 | 69.4 | 78.2 | 79.2 | 74 | 5.3 | 4 |
| 1905.5 | | 56.6 | 2.74 | 39.30 | 7.67 | | 56.6 | 51.1 | 61.9 | 52.2 | 55 | 4.9 | 4 |
| 1915.5 | | 65.0 | 3.21 | 40.15 | 8.62 | | 65.0 | 59.9 | 68.3 | 70.0 | 66 | 4.5 | 4 |
| 1925.5 | | 69.1 | 3.50 | 41.08 | 8.98 | | 69.1 | 65.3 | 75.4 | 76.7 | 72 | 5.4 | 4 |
| 1935.5 | | 87.6 | 4.47 | 42.64 | 9.45 | | 87.6 | 83.4 | 87.3 | 85.5 | 86 | 2.0 | 4 |
| 1945.5 | | 110.3 | 5.25 | 44.64 | 10.5 | | 110.3 | 97.9 | 102.6 | 105.2 | 104 | 5.2 | 4 |
| 1955.5 | | 132.6 | 6.28 | 46.81 | 11.09 | | 132.6 | 117.1 | 119.2 | 116.2 | 121 | 7.6 | 4 |
| 1965.5 | | 92.1 | 4.66 | 42.90 | 9.15 | | 92.1 | 86.9 | 89.3 | 79.9 | 87 | 5.2 | 4 |
| 1975.5 | | 99.4 | 5.30 | 43.41 | 9.62 | | 99.4 | 98.8 | 93.2 | 88.7 | 95 | 5.1 | 4 |
| 1985.5 | | 120.6 | 6.14 | 46.93 | 10.62 | | 120.6 | 114.5 | 120.1 | 107.4 | 116 | 6.1 | 4 |
| 1995.5 | | 101.2 | 5.27 | 44.50 | 10.08 | | 101.2 | 98.3 | 101.5 | 97.3 | 100 | 2.1 | 4 |
| 2005.5 | | 71.8 | 3.93 | 41.39 | 8.6 | | 71.8 | 73.3 | 77.8 | 69.6 | 73 | 3.5 | 4 |
| 2015.5 | | 50.0 | 3.30 | 39.48 | 7.73 | | 50.0 | 61.5 | 63.2 | 53.4 | 57 | 6.4 | 4 |

We can now put our Multimessenger reconstruction in the context of solar activity over the last millennium, Figure 3. It is encouraging that our reconstruction matches the WEA reconstruction very well ($R^2 = 0.87$) for their time of overlap, illustrating the power of the Multimessenger approach in reconciling various time series. We note that a ~100-year quasi-wave is clearly seen by eye over the last three centuries (only) and also that there has not been any significant secular change (e.g. an often claimed increase) over the same time interval, the lack of which had already been established (e.g. Clette et al. 2014).

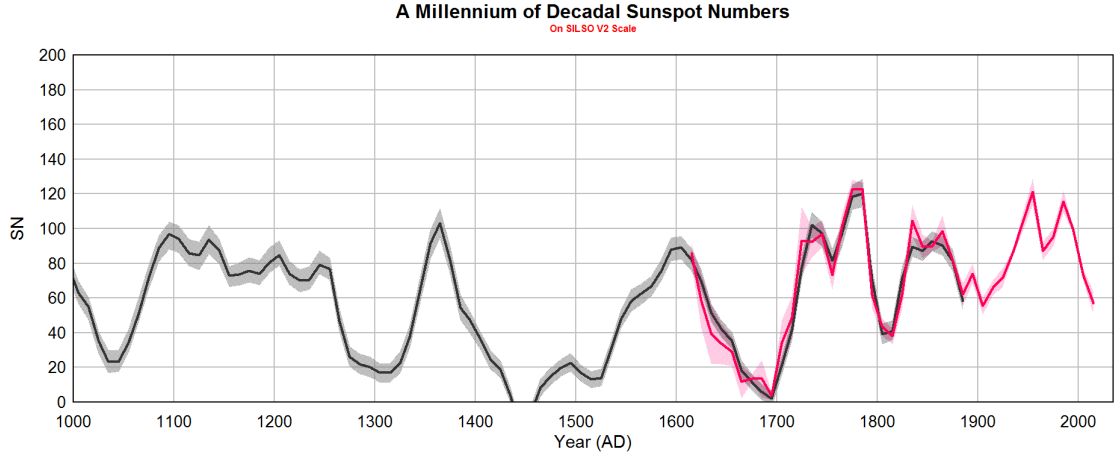

Figure 3. A millennium of reconstructed decadal sunspot numbers on the SILSO V2 scale. The WEA reconstruction is shown by the black curve with their stated uncertainty indicated by gray shading. The (unphysical) slightly negative values around 1450 AD have been omitted. They may be a hint that the reconstruction is problematic for very low activity where the usual assumption of a spherical symmetric solar wind breaks down. The average Multimessenger reconstruction for 1615-2015 AD is shown by the red curve with the uncertainty band indicated by pink shading. The two reconstructions generally agree within their uncertainties.

This convergence of the recent cosmogenic and solar activity records (see also Muscheler et al. 2016) lends credence to the admissibility of making a leap of faith back to the beginning of the WEA reconstruction nine millennia ago, Figure 4, even if we have to admit that it is not clear if the very long-period variations are of solar origin. On the other hand, it seems clear that recent activity has not been extraordinary (Berggren et al. 2009).

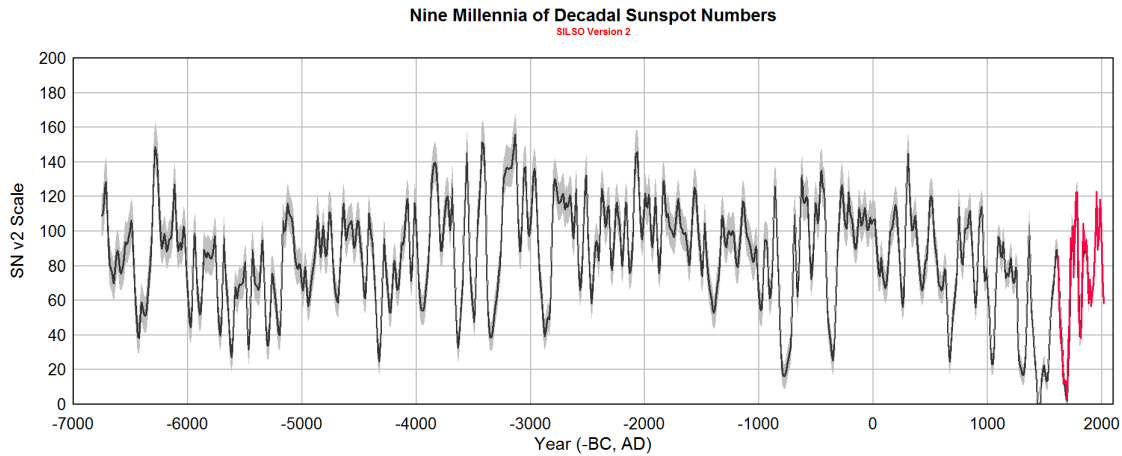

Figure 4. Nine millennia of reconstructed decadal sunspot numbers on the SILSO V2 scale. The WEA reconstruction is shown by the black curve with the stated uncertainty indicated by gray shading. The average Multimessenger reconstruction for 1615-2015 AD is shown by the red curve.

The combined time series from 6755 BC to 2015 AD is available as an Excel file at https://leif.org/research/Nine-Millennia-SN.xls.

**Periodic Activity?**

When you have 8770 years (878 data points) of data, the urge to look for cycles is overwhelming. Figure 5 shows the magnitude of the FFT in the time domain of the full sunspot number time series (combining the WEA and Multimessenger series). Although there are better and more powerful methods (e.g. Wavelets), any real periodic activity would show up in the FFT spectrum. We computed the FFT for the entire series with no windowing but with zero-appending as needed, and also for three subsets: the first half, the second half, and the middle half in order to see if periods ('cycles') would be persistent and coincident in all of them. Three long-term cycles are often assumed to exist (e.g. Damon & Sonett, 1991): the ~2300-year Hallstatt (or Bray) Cycle, the 208-year de Vries (or Suess) Cycle, and the 88-year Gleissberg Cycle. Figure 5 shows that the Hallstatt Cycle (found in climate records) is not significant in the solar record. There does seem to be power at periods between 200 and 240 years, but the power is perhaps too broadly distributed to qualify as a strong periodicity, although there is a narrow peak at half the period (104 years), a variation also visible by eye in Figure 3. With lots of peaks between 250 and 1200 years it is no surprise that some of them just coincide around 350 years. On the other hand, the 87.6-year Gleissberg peak is sharp and prevalent in the whole series and in all three sub-intervals, although being absent in recent data.

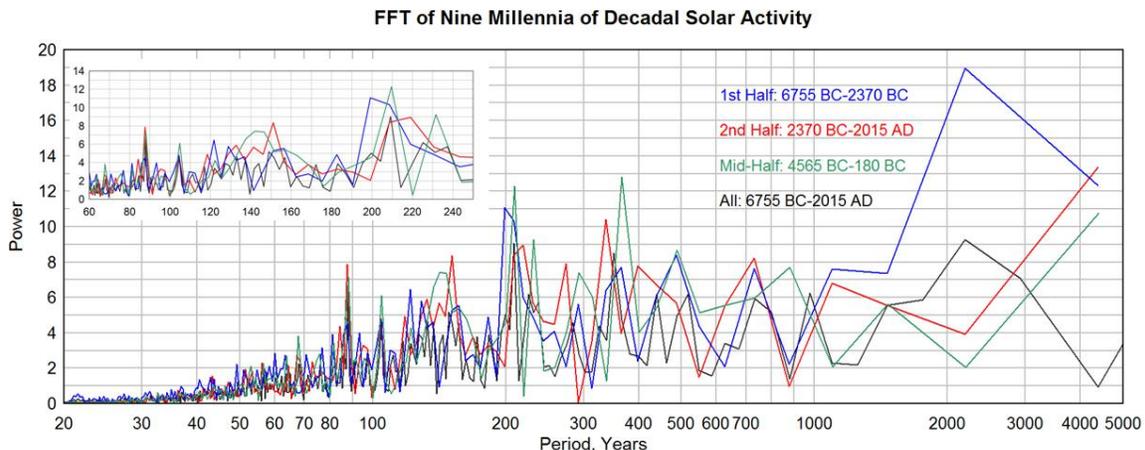

Figure 5. Fast Fourier Transform of the combined nine-millennia decadal sunspot number and of three subsets: first half, last half, and middle half. The insert shows the interval from 60 to 250 years in more detail.

**Conclusion**

The Wu et al. (2018) reconstruction of the sunspot number since 6755 BC combined with modern Multimessenger proxies for the last four hundred years goes a long way to reconcile the cosmogenic solar activity record with recent assessments of long-term solar activity.

**References**

Bartels, J. 1932, *Terr. Magn. Atmos. Electr.*, **37**, 1-52.
Berggren, A.-M., Beer J., Possnert, G., et al. 2009, *Geophys. Res. Lett.*, **36**(11), L11801.
Clette, F., Svalgaard, L., Vaquero, J. M., et al. 2014, *Space Sci. Rev.*, **186**, 35-103.
Clette, F., Lefèvre, L. 2016, *Solar Phys.*, **291**(9-10), 2629-2629.
Cliver, E. W., Herbst, K. 2018, *Space Sci. Rev.*, **214**(2), Id 56.
Damon, P. E., Sonett, C. P. 1991, in *The sun in time*, Univ. Arizona Press, 360-388.
Herbst, K., Muscheler, R., Heber, B. 2017, *J. Geophys.*, **122**(1), 23-34.
Loomis, E. 1873, *Amer. J. Sci. Ser. 3*, **5**, 245-260.
Muscheler, R., Adolphi, F., Herbst, K., et al. 2016, *Solar Phys.*, **291**(9-10), 3025-3043.
Owens, M. J., Cliver, E. W., McCracken, K. G., et al. 2016, *J. Geophys. Res.*, **121**(7), 6048-6063.
Suess, H. E. 1955, *Science*, **122** (3166), 415-417.
Svalgaard, L. 2014, *Ann. Geophysicae*, **32**(6), 633-641.
Svalgaard, L. 2016, *Solar Phys.*, **291**(9-10), 2981-3010.
Svalgaard, L., Cliver, E. W. 2005, *J. Geophys. Res.*, **110**(12), A12103.
Svalgaard, L., Cliver, E. W. 2010, *J. Geophys. Res.*, **115**(9), A09111.
Svalgaard, L., Schatten, K. H. 2016, *Solar Phys.*, **291**(9-10), 2653-2684.
Usoskin, I. G., Solanki, S. K., Kovaltsov, G. A. 2007, *A&A*, **471**, 301.
Usoskin, I. G., Solanki, S. K., Schüssler, M., et al. 2003, *Phys. Rev. Lett.*, **91**, 211101.
Wu, C. J., Usoskin I. G., Krivova, N., et al. 2018, *A&A*, **615**, A93.